# Characterizing and Modeling Control-Plane Traffic for Mobile Core Network

JIAYI MENG, Purdue University, USA
JINGQI HUANG, Purdue University, USA
Y. CHARLIE HU, Purdue University, USA
YARON KORAL, AT&T Labs, USA
XIAOJUN LIN, Purdue University, USA
MUHAMMAD SHAHBAZ, Purdue University, USA
ABHIGYAN SHARMA, AT&T Labs, USA

In this paper, we first carry out to our knowledge the first in-depth characterization of control-plane traffic, using a real-world control-plane trace for 37,325 UEs sampled at a real-world LTE Mobile Core Network (MCN). Our analysis shows that control events exhibit significant diversity in device types and time-of-day among UEs. Second, we study whether traditional probability distributions that have been widely adopted for modeling Internet traffic can model the control-plane traffic originated from individual UEs. Our analysis shows that the inter-arrival time of the control events as well as the sojourn time in the UE states of EMM and ECM for the cellular network cannot be modeled as Poisson processes or other traditional probability distributions. We further show that the reasons that these models fail to capture the control-plane traffic are due to its higher burstiness and longer tails in the cumulative distribution than the traditional models. Third, we propose a two-level hierarchical state-machine-based traffic model for UE clusters derived from our adaptive clustering scheme based on the Semi-Markov Model to capture key characteristics of mobile network control-plane traffic—in particular, the dependence among events generated by each UE, and the diversity in device types and time-of-day among UEs. Finally, we show how our model can be easily adjusted from LTE to 5G to support modeling 5G control-plane traffic, when the sizable control-plane trace for 5G UEs becomes available to train the adjusted model. The developed control-plane traffic generator for LTE/5G networks is open-sourced to the research community to support high-performance MCN architecture design R&D.

## 1 INTRODUCTION

The Mobile Core Network (MCN) is at the heart of the cellular network; it manages and tracks all users' activity (including mobility) as well as forwards data traffic between users and the Internet. To efficiently and flexibly handle control and data traffic in these networks, Control-/User-Plane Separation (CUPS) was first introduced in 3GPP Release 14 for 4G [1] and further refined in 3GPP Release 15 for 5G [4], to separate cellular operations between a control plane (for managing control traffic) and a data plane (for forwarding data traffic).

In the past few years, as 5G deployment has been gaining momentum, cellular networks have witnessed not only an explosive growth in data-plane traffic, but also a significant increase in control-plane traffic. For example, the control-plane traffic has been reported to grow 50% faster than data-plane traffic since 2015 [35], and some carriers are reportedly experiencing more than 100× increase in the volume of transactions in the 5G control plane compared to 4G in 2021 [7]. Such continuous traffic growth challenges mobile network operators and designers to innovate on mobile network architectural design and performance optimization, constantly. Given the clear separation of the control plane and the data plane, it is essential to innovate on both planes of MCN to ensure a high-quality mobile user experience.

However, the large body of recent work on LTE and 5G has focused on the data plane of MCN, i.e., on modeling data-plane traffic (e.g., [13, 15, 22, 27–29]) and improving data-plane performance and scalability [10, 23, 24, 30, 34, 36, 38]. Several recent work has studied the control plane of MCN, but has primarily focused on improving control-plane performance without a clear understanding of control-plane traffic [9, 23, 24, 34, 38]. Therefore, it remains unclear how well their designs will perform with realistic large-scale control-plane traffic today as well as tomorrow for NextG.

We argue that for MCN to sustain the rapid growth in traffic demand in the coming years, it is equally vital and urgent to study the impact of control-plane traffic on the MCN performance and consequent design improvement, and conducting such studies critically relies on high-fidelity large-scale control-plane traffic to drive the MCN in order to evaluate and validate MCN designs under realistic workloads.

Despite the need for large-scale control-plane traffic in MCN, such data is only accessible by mobile network operators, who are reluctant to directly share their traffic traces due to business and privacy concerns. As a result, the lack of public MCN control-plane traffic hinders the in-depth study of MCN design and performance optimization by the broad networking and systems communities. While previous work [15] studied the control-plane traffic, it has mostly focused on theoretical traffic modeling without looking into real-world traffic behavior.

**Our Contributions.** This paper makes four contributions toward modeling and developing a readily-usable event generator for the control-plane traffic of LTE/5G mobile networks.

• **First,** we perform to our knowledge the first in-depth characterization study of the control-plane traffic of an LTE network from one major carrier in the US by analyzing a 7-day trace containing a total of 196, 827, 464 control-plane events due to 37, 325 real UEs—consisting of three primary types of devices, i.e., phones, connected cars, and tablets. We choose to study an LTE trace, because 5G deployment is still at an initial stage and systematic extensive trace collection is not yet available.

Our analysis reveals that the control-plane events exhibit significant diversity in device types and time-of-day among UEs: (1) The percentages of Service Request (SRV_REQ) and S1 Connection Release (S1_CONN_REL) are 5.1%~17.0% higher for phones and tablets than for connected cars, while the percentages of Handover (HO) and Tracking Area Update (TAU) events are 74.2%~212.4% higher for connected cars than for phones and tablets. (2) The average per device-hour volume of control events for the four dominant control-plane events (i.e., SRV_REQ, S1_CONN_REL, HO, and TAU) drops significantly from the peak hour to the slowest hour of the day by 2.27×, 2.28×, 86.15×, and 6.25× for phones, by 3.48×, 3.43×, 1309.33×, and 3.89× for connected cars, and by 1.48×, 1.45×, 90.06×, and 2.11× for tablets, respectively. (3) Among UEs of the same device type in the same hour, the differences between maximum and minimum numbers of those four dominant event types are also large across different hours of the day, i.e., 108~138, 112~142, 2~43, and 4~24 for phones; 18~98, 24~105, 1~36, and 4~19 for connected cars; and 124~169, 128~175, 0~23, and 2~13 for tablets.

• **Second,** we study whether traditional probability distributions that have been widely used for modeling Internet traffic can be readily used to model the control traffic originated by individual UEs in the mobile network. Building on the insights about the diversity of control-plane traffic, we first divide up the traffic into sub-traces, one for each distinct combination of UE cluster (segregated using our proposed adaptive clustering scheme), 1-hour interval and device type, and perform two standard statistical tests, the Kolmogorov-Smirnov test [32] and the Anderson-Darling test [40], on each subtrace. Our statistical test results show that surprisingly, the inter-arrival time of the control events and the sojourn time in the four UE states of EMM and ECM (DEREGISTERED, REGISTERED, CONNECTED, and IDLE) for the mobile network cannot be modeled as Poisson processes or other traditional probability distributions, including Pareto [21], Weibull [39] and Tcplib [16, 17].

We further study the reasons why these traditional probability distributions fail to model control traffic by analyzing how well the Poisson distribution can model the burstiness of control-plane traffic via variance-time plots [19, 26], and directly comparing the cumulative distributions of the trace with the fitted Poisson distributions. Our analysis reveals that the control-plane traffic of the mobile network has much higher burstiness and longer tails, in their cumulative distributions, compared to the traditional probability models.

• **Third,** we propose a two-level hierarchical state-machine-based traffic model for each UE cluster derived from our adaptive clustering scheme based on the Semi-Markov Model to capture the traffic diversity as well as the dependence among events generated by each UE required by the 3GPP.

To properly capture the traffic diversity among the UEs, the straightforward design choice is to first cluster the traffic for each event type, and then derive a traffic model per cluster for every event type. However, such an approach ignores the dependence among different types of events and thus the dependence among the models derived for different event types. To address this, we develop an adaptive clustering scheme to recursively cluster the UEs based on their traffic similarity for each 1-hour interval and every device type, and then derive models for all event types for each UE-cluster/hour/device-type combination. Such an approach enables us to capture both the dependence among different types of events and the diversity in those three dimensions.

To capture the dependence among events for every single UE (following the 3GPP protocol [2]), we instantiate the two-level state-machine-based model (i.e., sojourn time in states and state transition probabilities) for each UE-cluster/hour/device-type. Specifically, we observe that the EMM/ECM state machines defined by the 3GPP can accurately capture the four types of events (ATCH, DTCH, SRV_REQ, and S1_CONN_REL) that trigger the UE state transitions, but not the two types of events (HO and TAU) that do not change the UE states. We therefore introduce two fine-grained sub-state machines that are embedded inside CONNECTED and IDLE states of the EMM–ECM state machine to capture the intricate dependence of HO and TAU events on other events. We further adopt the Semi-Markov model [47] to model the duration of a UE staying in the current state and the probability of transitioning to the next state. Unlike the Markov model, the Semi-Markov model does not assume that the sojourn time in the same state follows the exponential distribution with a constant hazard rate, which as we show is not applicable to mobile network control traffic.

To experimentally validate our state-machine-based traffic model, we develop a baseline method using the fitted Poisson distributions and compare the synthesized traces using the baseline and our method against two fresh testing real traces for 38K and 380K UEs, respectively. We show that our method outperforms the baseline from both macroscopic and microscopic perspectives: (1) We first compare the breakdown of the synthesized events with that of the real events. Compared with the real traces, our synthesized traces achieve small differences, i.e., within 1.7%, 5.0% and 0.8%, for phones, connected cars, and tablets, respectively, while the traces generated by the baseline have the differences up to 37.2%, 38.5% and 33.0%, for both UE population sizes. (2) We then compare the per-UE traffic behavior, including the numbers of events per UE and the sojourn time in the UE states for the two dominant state transitions (i.e., between CONNECTED and IDLE). Compared with the baseline, for both test traces and for phones, our method reduces the maximum y-distance of the CDF of events per UE between the synthesized and actual traces by over 7.74× and 7.46× for SRV_REQ and S1_CONN_REL events, and the maximum y-distance of the CDF of the sojourn time in CONNECTED and IDLE states between the synthesized and actual traces by over 4.77× and 3.25×. Similar improvements are observed for connected cars, by over 1.15×, 1.18×, 2.65×, and 1.23×, and for tablets, by over 3.17×, 3.05×, 8.56×, and 2.80×. These results suggest our model can synthesize traces for much larger number of UEs than the trace we use to instantiate the model.

• **Lastly,** since more and more UEs are migrating to 5G from earlier generations of mobile networks, we present a methodology to showcase how easy it is to adjust our proposed traffic model for LTE to for 5G; the parameters of the new traffic model for 5G can be readily seeded with a large-scale control-plane trace for 5G UEs when it becomes available, or directly scaled from the 4G model.

We open-source the developed control-plane traffic model to the community to stimulate further research on MCN design and optimization, not only for 4G, but also for 5G and beyond. [1]

---

[1]Contact authors for downloading information.

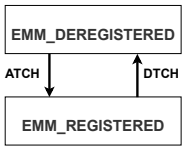 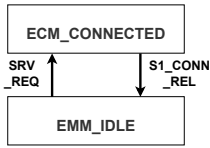 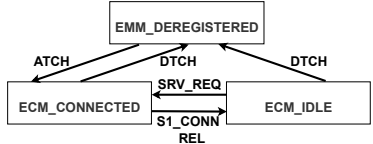

Fig. 1. EMM state machine.  Fig. 2. ECM state machine.  Fig. 3. EMM−ECM state machine.

Table 1. Breakdown of control-plane events for different types of devices in the 7-day trace.

| Event | Event Type | Phones | Connected Cars | Tablets |
| --- | --- | --- | --- | --- |
| Attach [2] | ATCH | 0.1% | 0.9% | 1.2% |
| Detach [2] | DTCH | 0.2% | 0.9% | 1.1% |
| Service Request [2] | SRV_REQ | 45.5% | 38.9% | 43.9% |
| S1 Connection Release [2] | S1_CONN_REL | 47.5% | 45.2% | 47.7% |
| Handover [2] | HO | 3.8% | 6.6% | 2.1% |
| Tracking Area Update [2] | TAU | 2.9% | 7.4% | 4.0% |

## 2 BACKGROUND

We begin with a brief background of LTE network architecture and its control-plane events, and review LTE traffic modeling work in the literature.

### 2.1 LTE Network Architecture

The LTE mobile network consists of three components: User Equipment (UE), Radio Access Network (RAN) and Evolved Packet Core (EPC), which finally connects to the Internet to provide data services to UEs. A UE is a device used by an end-user to communicate with the mobile network (e.g., phone, tablet, IoT, etc.). RAN resides between UEs and EPC and manages the radio spectrum of a distributed collection of base stations that directly communicate with UEs via wireless signals.

EPC represents the MCN of LTE. It consists of the following five network functions: (1) *Mobility Management Entity* (MME): It provides mobility session management for the LTE network and supports subscriber authentication, roaming, and handovers; (2) *Home Subscriber Server* (HSS): It is a central user database that stores subscription-related information; (3) *Policy and Charging Rules Function* (PCRF): It tracks and manages policy rules, and also logs data charges on subscriber traffic; (4) *Packet Data Network Gateway* (PGW): It provides connectivity from a UE to the Internet by being its point of exit and entry of data traffic; and (5) *Serving Gateway* (SGW): It acts as a router and forwards data packets between RAN and PGW.

To efficiently and flexibly handle control and data traffic, EPC is partitioned into a control plane and a user plane [2]. The control plane manages signaling traffic between RAN and MME and the other network functions of MCN (i.e., HSS, PCRF, SGW, and PGW). The user plane forwards data traffic among RAN, SGW and PGW, and thus is also called the data plane.

### 2.2 LTE Control-Plane Events

Table 1 summarizes major LTE control-plane events between UE and MCN.[2] In the control plane [2], (1) *Attach* (ATCH) registers UE with MCN; (2) *Detach* (DTCH) deregisters UE from MCN, when the UE is switched off; (3) *Service Request* (SRV_REQ) creates a signaling connection for the UE to send/receive signaling messages or data; (4) *S1 Connection Release* (S1_CONN_REL) releases the signaling connection in the control plane, and other resources associated with the UE in the data plane; (5) *Handover* (HO) switches the UE from current serving cell to another cell; (6) *Tracking Area Update* (TAU) updates UE's tracking area, when the UE moves to another tracking area (comprising a new set of cells) or the periodic timer of TAU is expired or some other cases, e.g., the UE reselects to LTE from 3G, the UE re-registers to LTE after the fallback for circuit-switched services, etc.

---

[2]As we focus on the control plane of MCN in this paper, we ignore events that happen only between UE and RAN.

**Dependence among events.** The control events listed in Table 1 generated by a UE are not independent, as required to conform to the 3GPP protocol. The protocol specifies that one UE follows two independent state machines when interacting with MCN: *EPS Mobility Management* (EMM) and *EPS Connection Management* (ECM) [2].[3] Some control-plane events, such as ATCH, DTCH, SRV_REQ, and S1_CONN_REL, can trigger changes to the UE states, while others can not change the UE states but still have intricate dependence on those states and thus the corresponding events.

(1) The *EMM state machine* describes the UE's Mobility Management states that maintain the information related to UE's registration with the MCN. Figure 1 shows the two primary EMM states, EMM_DEREGISTERED and EMM_REGISTERED (denoted as DEREGISTERED and REGISTERED), and the corresponding control events that trigger transitions between them. Generally, the UE can be powered on (off), which in turn triggers ATCH (DTCH) event to enter the REGISTERED (DEREGISTERED) state.

(2) The *ECM state machine* describes the signaling connectivity between the UE and the MCN, when the UE stays in REGISTERED. Figure 2 shows the two primary ECM states, ECM_CONNECTED and ECM_IDLE (denoted as CONNECTED and IDLE), and the corresponding control events that trigger state transitions. Generally, SRV_REQ (S1_CONN_REL) can be triggered to switch the UE state to CONNECTED (IDLE). For the other control-plane events, some (e.g., TAU) can happen in both CONNECTED and IDLE, while the rest (e.g., HO) can only happen in CONNECTED.

## 2.3 Prior LTE Traffic Modeling

Most of previous work on traffic modeling for LTE networks [13, 15, 22, 27–29, 31, 33, 42–44, 48, 49] focuses on modeling the data-plane traffic, instead of control-plane traffic. For LTE control-plane traffic, Dababneh et al. [15] proposed to model the total control-plane volume on different functions of LTE's MCN, using the number of UEs and events' transactions per second per subscriber. However, they ignored the traffic diversity in device types and time-of-day, and they also did not model the fine-grained inter-arrival time of successive events for individual UEs.

## 3 CHARACTERIZING REAL-WORLD LTE CONTROL-PLANE TRAFFIC

In this section, we characterize the LTE control-plane events generated by an extensive collection of real UEs of a major mobile carrier in the US.

**Dataset.** We sampled 37,325 UEs and collected their control-plane events recorded from MMEs over one whole week of June in 2022. The timestamps collected have a millisecond granularity. In total, we collected 196,827,464 events.

### 3.1 Diversity in Device Types

To analyze the diversity of control-plane traffic across different types of devices, we first calculate numbers of the sampled UEs for different device types, and then explore the distributions of various types of control-plane events for these devices as well as the duration of devices staying in the four EMM/ECM states caused by the control-plane events.

**Breakdown of UEs.** We first categorize the UEs into three primary types of devices: *phones, connected cars*, and *tablets*. We derive the device type of every UE via the Type Allocation Code (TAC), which is the first eight digits of UE's IMEI, and can identify the corresponding manufacturer and device model and thus the device type [46]. Of all sampled UEs, 23,388 are phones, 9,308 are connected cars, and 4,629 are tablets.

**Breakdown of control events.** Table 1 shows that different device types have diverse distributions of control-plane events. For phones, SRV_REQ and S1_CONN_REL dominate the total control-plane events at 45.5% and 47.5%, respectively, HO accounts for 3.8% of the total events, and TAU accounts for 2.9% of the total events. This highly skewed distribution shows that phone users have a higher frequency of sending/receiving data to/from the Internet; they tend not to switch on/off phones often, to allow the UEs to access network services at all times.

---
[3]EPS stands for Evolved Packet System, which comprises RAN and EPC.

Table 2. Breakdown of duration of staying in the EMM and ECM states for different device types.

| | State | Phones | Connected Cars | Tablets |
|---|---|---|---|---|
| EMM | REGISTERED | 98.1% | 97.3% | 90.0% |
| | DEREGISTERED | 1.9% | 2.7% | 10.0% |
| ECM | CONNECTED | 16.1% | 7.2% | 13.1% |
| | IDLE | 82.0% | 90.0% | 76.8% |

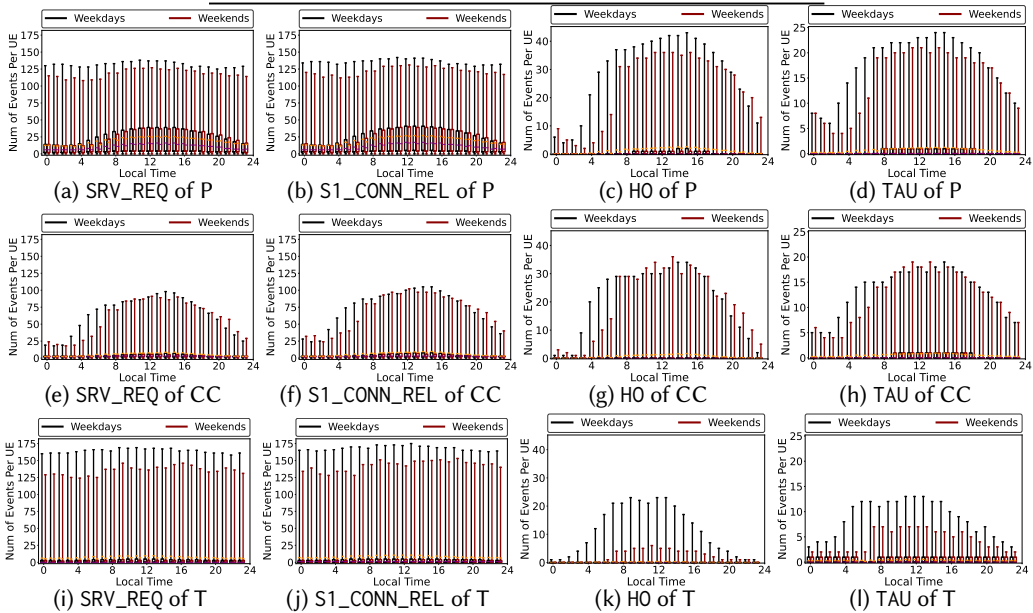

Fig. 4. Box plots of numbers of control events per device-hour of different types of devices over 24 hours.

For connected cars, the percentages of SRV_REQ and S1_CONN_REL are 38.9% and 45.2%, respectively; both lower than those of phones, while the percentages of HO, TAU, ATCH and DTCH (6.6%, 7.4%, 0.9%, and 0.9%, respectively) are higher than those of phones. This disparity suggests that connected cars have different behavior from phone users: the higher mobility of cars incurs a larger volume of HO; the vehicle-to-everything technology activated by additional IoT sensors (e.g., LiDAR, camera) installed on connected cars further triggers more ATCH and DTCH than phones.

Unlike phones and connected cars, tablets have only 2.1% and 4.0% of the total events for HO and TAU, respectively, as users tend to remain stationary while using tablets. Tablets have similar percentages of ATCH and DTCH (1.2% and 1.1%, respectively) as connected cars, higher than those for phones, as users likely turn on/off tablets more frequently than phones to save battery.

**Breakdown of duration in EMM/ECM states.** Table 2 shows that different types of devices spend varying duration in the EMM and ECM states. For EMM, phones and connected cars spend similar time in REGISTERED, i.e., 98.1% and 97.3%, respectively. Compared with phones and connected cars, tablets spend less time in REGISTERED, i.e., 90.0%, as users likely turn on/off their tablets to save battery. For ECM, phones and tablets spend similar percentages of time in CONNECTED to provide data services, i.e., 16.1% and 13.1%, respectively. Compared with phones and tablets, connected cars are much less active, spending 7.2% of time in CONNECTED.

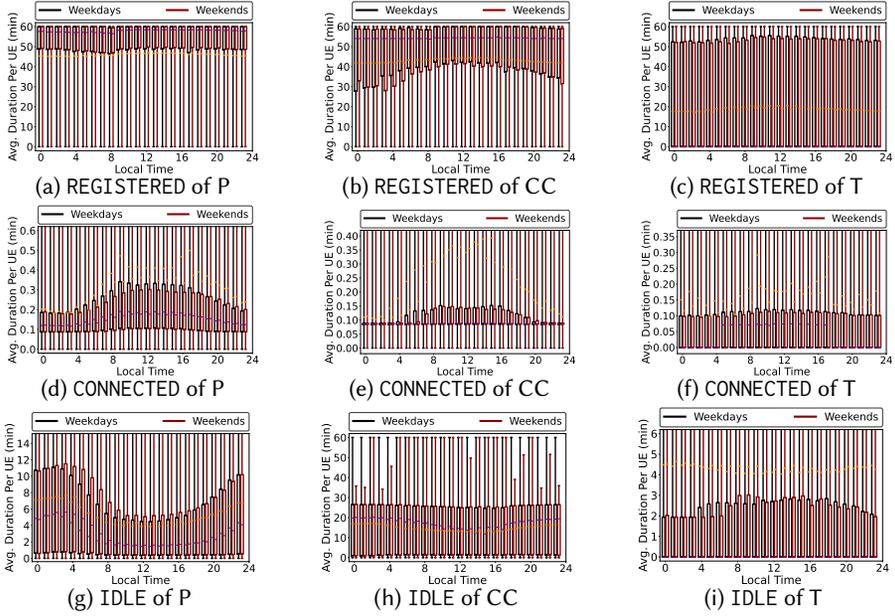

Fig. 5. Box plots of average duration of staying in REGISTERED, CONNECTED and IDLE for three types of devices over 24 hours. The maximum average duration per UE is 60 minutes for all, except IDLE of connected cars.

## 3.2 Diversity in the Time-of-Day among UEs

To understand how the control-plane traffic is affected by the time-of-day among UEs, we divide the trace into 1-hour intervals for each day. We analyze the diversity of numbers of events per UE as well as average duration of each UE staying in different EMM/ECM states across different hours of the day, using box plots. The details on how we plot them are in Appendix A. We then zoom into TAU that can occur in both CONNECTED and IDLE, and examine how the time-of-day affects the percentages of TAU in CONNECTED and IDLE among UEs. Meanwhile, we also consider weekdays and weekends separately to study whether the characteristics of the control-plane traffic differ.

**Number of events per UE.** Figure 4 shows similar diurnal patterns across different types of devices for the four dominant control-plane event types (SRV_REQ, S1_CONN_REL, HO, and TAU), during both weekdays and the weekend. For SRV_REQ, during weekdays, phones start with an average frequency of 12.59 events per device-hour at around 12 AM, reach the lowest frequency of 11.73 at 2 AM, and then increase to the highest frequency of 25.86 at around 1 PM, as the users are getting increasingly active. The users then gradually wind down and the frequency of SRV_REQ gradually decreases back to 13.67 per device-hour at the lowest at 11 PM. During the weekend, the average frequency of phones varies between 10.62 at 3 AM and 24.80 at 12 PM.

Compared to phones, connected cars have about 73.2% (weekdays) and 74.1% (weekends) fewer SRV_REQ events per device-hour; it starts from 2.35 (weekdays) and 2.43 (weekends) at 12 AM, increases to 8.65 at 3 PM (weekdays) and 7.44 at 11 AM (weekends), and finally drops to 2.50 (weekdays) and 2.57 (weekends) at 11 PM. Tablets also have small variations in the SRV_REQ frequency over different hours, as the frequency per device-hour peaks at 9.62 at 1 PM (weekdays) and 6.75 at 11 AM (weekends) and settle at 6.15 at 12 AM (weekdays) and 4.86 at 3 AM (weekends).

As with SRV_REQ, S1_CONN_REL has slightly higher frequency for all three device types over all hours of the day, i.e., 12.18~26.99 (weekdays) and 11.07~25.95 (weekends) for phones, 2.66~9.68 (weekdays) and 2.62~8.40 (weekends) for connected cars, and 6.91~10.63 (weekdays) and 5.56~7.63 (weekends) for tablets. This is because most of S1_CONN_REL events are triggered to release the signaling connections created by SRV_REQ earlier.

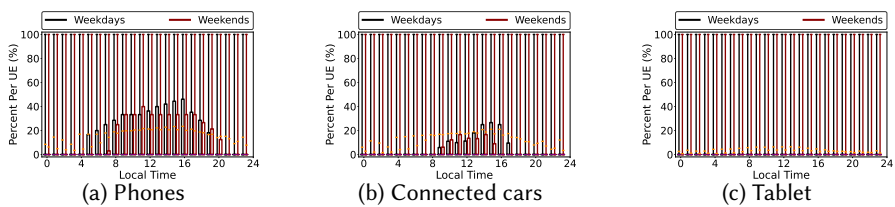

(a) Phones  (b) Connected cars  (c) Tablet

Fig. 6. Box plots of percentages of TAU in CONNECTED per UE for different types of devices over 24 hours.

Similarly, HO also follows diurnal patterns during weekdays and the weekend for three types of devices. For example, phones start with 0.07 (weekdays) and 0.11 (weekends) events per device-hour at 12 AM. Then the average frequency increases to 2.63 at 3 PM (weekdays) and 2.20 at 12 PM (weekends), since users are becoming increasingly active. Finally, the frequency drops back to 0.15 (weekdays) and 0.22 (weekends) at 11 PM, which shows their much lower mobility at night.

In contrast with HO, the frequency of TAU has smaller variations over different hours of the day for all three types of devices: 0.23~1.25 (weekdays) and 0.16~1.14 (weekends) for phones, 0.25~0.94 (weekdays) and 0.21~0.84 (weekends) for connected cars, and 0.29~0.54 (weekdays) and 0.16~0.37 (weekends) for tablets. Such small variations are because TAU can be triggered, when the periodic timer for TAU expires or the UE moves to a new tracking area that covers another set of base stations. Therefore, TAU is less frequent than HO.

In addition to the diurnal patterns, we also observe highly diverse numbers of events among UEs of the same device type every single hour. Specifically, the average differences between maximum and minimum numbers of SRV_REQ per UE across different hours of the day are 124.96, 61.63 and 149.67 for phones, connected cars, and tablets, respectively. Compared with SRV_REQ, S1_CONN_REL has slightly higher average differences, i.e., 128.90, 71.21 and 155.38 for the three types of devices. As for HO and TAU, which happen much less frequently than SRV_REQ and S1_CONN_REL, the average differences between maximum and minimum per-UE frequencies for three types of devices are still 26.27, 20.35 and 7 for HO, and 16.17, 12.81 and 6.52 for TAU.

**Time spent in EMM/ECM states.** For the EMM states, Figure 5 (top) shows that the duration spent in REGISTERED does not vary much over the time-of-day for all three types of devices. Specifically, phones spend 45.17~46.83 minutes (weekdays) and 45.06~46.91 minutes (weekends) in REGISTERED, and connected cars spend 41.85~44.54 minutes (weekdays) and 41.83~44.12 minutes (weekends) in REGISTERED on average. Compared with phones and connected cars, tablets spend 55.7%~58.4% less time in REGISTERED, i.e., 17.64~20.48 minutes (weekdays) and 17.24~19.43 minutes (weekends) on average. Regardless of the time of the day, the duration in REGISTERED still exhibits high diversity (i.e., 0~60 minutes) among UEs of each device type in every single hour, as there are always UEs that are not powered on across different hours of the day.

As for the ECM states, Figure 5 (middle) shows the average duration of CONNECTED sessions per UE across different time-of-day for different types of devices. [4] For phones, the average duration starts from 0.20 minutes at 12 AM (weekdays and weekends), increases to 0.50 minutes (weekdays) and 0.37 minutes (weekends) at 5 PM, and drops back to 0.23 minutes (weekdays) and 0.24 (weekends) at 11 PM. The average duration is 0.11~0.59 minutes (weekdays) and 0.10~0.37 minutes (weekends) for connected cars, and 0.13~0.29 minutes (weekdays) and 0.10~0.29 minutes (weekends) for tablets. Moreover, the duration of CONNECTED is also diverse among UEs of the same device type across all hours for all three types of devices. The maximum duration per CONNECTED is 60 minutes, 15.07~60 minutes and 30.06~60 minutes for phones, connected cars, and tablets, respectively.

---

[4]There can be multiple CONNECTED and IDLE sessions within a single hour for each UE.

Similarly, for IDLE, Figure 5 (bottom) presents the diversity across the time-of-day during weekdays and the weekend (i.e., 3.86~7.69 minutes for phones, 12.64~17.09 minutes for connected cars and 4.03~4.65 minutes for tablets), as well as the diversity among UEs of the same device type in each hour ranging from 0 to 60 minutes for all three types of devices.

**Breakdown of TAU by ECM states.** Considering TAU can occur in both CONNECTED and IDLE, we next analyze how TAU is affected by those states among UEs across different hours of the day.

Figure 6 shows that the percentage of TAU in CONNECTED per UE not only varies across different hours of the day, but also exhibits high diversity among UEs for each device type. The average percentages of TAU in CONNECTED per UE are 8.52%~23.48% (weekdays) and 3.44%~21.64% (weekends) for phones, 4.42%~21.06% (weekdays) and 1.59%~18.80% (weekends) for connected cars, and 1.89%~6.43% (weekdays) and 0.66%~3.78% (weekends) for tablets. The maximum percentages reach 100% across all hours of the day for all three types of devices.

## 4 MODELING CONTROL-PLANE TRAFFIC OF CELLULAR NETWORKS

We next discuss the motivation and the design goals for modeling and generating control traffic for cellular networks, as well as the rationale for modeling *individual* UE's control-plane traffic.

**Motivation.** With 5G deployment gaining momentum as well as the number of 5G-connected devices skyrocketing, the volume of control-plane traffic is also escalating. For example, several cellular carriers are already experiencing more than 100× increase in the 5G control plane transaction volume compared to 4G in 2021 [7]. This significant traffic growth motivates the need for continuous innovation in mobile network architectural design and performance optimization. Enabling such architectural design innovations critically relies on developing accurate, scalable and versatile control traffic generators, in order to precisely profile and debug the mobile network performance, not only for realistic traffic load today but also in the (near) future, as more UEs migrate to 5G networks. Despite the large body of work on data-plane traffic modeling and generation, there has been little work on control-plane traffic modeling for mobile networks.

**Design goals.** The mobile network's control traffic generator for supporting the above usage must meet the following requirements:

(1) **Accuracy:** The generator outputs realistic control-plane traffic for a fixed UE population, i.e., it can accurately capture the inter-arrival duration of each type of event.
(2) **Event-Owner Labeling:** Every event in the generated control traffic is labeled with its originating UE. This is required to properly drive the network functions of the MCN, e.g., a UE transition among EMM and ECM states in both 4G and 5G.
(3) **Scalability:** The generator outputs realistic control traffic for an arbitrary UE population, e.g., for evaluating the scalability of an MCN design under increasing workload;
(4) **NextG Network Support:** The generator must support next-generation cellular networks, e.g., generating realistic control events from 5G UEs.

Similarly, like modeling Internet traffic, modeling control-plane traffic also needs to model packet (i.e., control event) inter-arrival time. Contrary to modeling Internet traffic, modeling control-plane traffic does not need to model the sizes of the control events, because following the 3GPP specification each control event has a fixed and small size.

**Why not modeling aggregate control-plane traffic?** We envision the primary use of the control traffic generator is for evaluating the performance of an MCN design and, hence, should generate the aggregate control traffic observed by the MCN. As such, we could try to directly model the aggregate control traffic for a given UE population, by fitting the aggregate traffic in our trace collection using some well-known probabilistic distributions, as in the prior work for modeling

Internet traffic, and generate synthesized traffic using the fitted distributions. However, such an approach has two limitations: (1) It is oblivious to, and cannot capture, the inter-dependence of control events of individual UEs dictated by the EMM and ECM state machines (§2.2). In particular, it will not be able to label each control event generated with a proper UE id, which is needed to correctly drive event processing performed by MCN functions. (2) Since such a model is derived by fitting a control traffic trace on a fixed UE population, it is often difficult to generate traffic for different UE population sizes, e.g., if the fitted model is Pareto distribution or TCPlib distribution. We, thus, propose to develop a methodology that models the control traffic of individual UEs, which can then be used to build a traffic generator that meets all the design requirements discussed above.

In what follows, we first study how to model the individual UE traffic by answering the following three questions: (1) Can traditional probability distributions, which are widely adopted for Internet traffic, model the individual UE traffic of LTE's control plane? If not, (2) why do those distributions fail? And, (3) how to model the individual UE traffic? We finally test the developed individual traffic models and the generator under two validation scenarios.

### 4.1 Can Traditional Probability Distributions Model Individual UE Traffic?

Network traffic modeling has been a foundation area of research over the history of the Internet. Many probability distributions have been leveraged to model the inter-arrival time of Internet traffic. We summarize these Internet traffic models as follows.

(1) The *Poisson distribution* is the predominant model used for modeling network traffic arrivals [11, 18, 25, 37]. It has been proven valid for modeling the arrival time of user-initiated sessions in wide-area networks, e.g., TELNET connections and FTP control connections [37]. Specifically, a Poisson process characterizes the inter-arrival duration $D_n$ as independently and exponentially distributed with a fixed rate parameter $\lambda$, i.e., $P(D_n > t) = e^{-\lambda t}$.
(2) The *Pareto distribution* [21] has been applied to model self-similarity in packet traffic of the wide-area network [6, 18]. It models the inter-arrival time by a power-law probability distribution that follows the probability density function of $f(x) = \alpha x_m^\alpha x^{-(\alpha+1)}$, where $\alpha$ is the shape parameter and $x_m$ is the minimum possible value of $x$ (normally, $x_m = 1$) [21].
(3) The *Weibull distribution* [39] has been shown to capture the inter-arrival time dynamics from different Internet traffic levels (sessions, flows and packets) [8]. A Weibull process follows the probability density function of $f(x) = \frac{k}{\lambda} (\frac{x}{\lambda})^{k-1} e^{-(x/\lambda)^k}$ for $x \geq 0$, where $k$ is the shape parameter and $\lambda$ is the scale parameter.
(4) The *TCPlib distribution* is an empirical distribution proposed to model the inter-arrival time within TELNET connections of real wide-area TCP/IP traffic [18].

However, whether those theoretical distributions can model LTE's control-plane traffic of individual UEs is unclear. We next answer this question by addressing two following challenges: (1) How do traditional probability distributions capture the diversity in device types and time-of-day among UEs? (2) How to validate if the traffic can be modeled using those distributions?

#### 4.1.1 Applying traditional probability distributions.
Considering the diversity in the control-plane traffic in terms of device types and time-of-day among UEs (shown in §3), we preprocess the input control traffic trace by dividing the entire trace into non-overlapping 1-hour intervals. Within each interval, we calculate the inter-arrival time for every type of control events for each UE. However, each UE has a limited number of events per hour. It is not practical to examine every single UE for every hour. Assuming the events of different UEs are i.i.d., a straightforward workaround is to pool together and analyze the inter-event arrival time of many UEs of the same device type for each hour. Note that it is not the same as merging the traces of many UEs into a *single* trace, which would change the inter-arrival time for each event. Still, simply pooling the inter-arrival time of all UEs does not work well, because the UEs of the same device type have diverse control-plane traffic (shown in §3). Thus, instead of polling the inter-arrival time from all UEs, we cluster the UEs of the same device type using the proposed adaptive clustering scheme (discussed later in §4.3), so that the UEs in the same cluster have similar traffic characteristics. We combine the inter-arrival time for all the UEs in the same cluster for every type of control event for every hour. Considering there exist repetitive diurnal patterns of control traffic across different days (Fig. 4 and Fig. 5), we then group the inter-arrival time of the same hour from different days together for each UE-cluster/hour/device-type combination. Finally, we fit each combination with traditional probability models using Maximum Likelihood Estimation (MLE) for each event type.

Table 3. Percentages of the 1-hour intervals whose inter-arrival time of different event types or sojourn time in the four EMM/ECM states passes the two standard statistical tests for the Poisson distribution.

| Test | Device Type | ATCH | DTCH | SRV_REQ | S1_CONN_REL | HO | TAU | REG. | DEREG. | CONN. | IDLE |
|---|---|---|---|---|---|---|---|---|---|---|---|
| K–S | Phones | 2.0% | 5.0% | 0.5% | 0.5% | 0.1% | 0.0% | 0.0% | 0.0% | 0.0% | 0.2% |
| | Conn. Cars | 2.5% | 5.0% | 0.2% | 0.0% | 0.0% | 0.0% | 0.0% | 0.0% | 0.0% | 0.0% |
| | Tablets | 0.0% | 0.6% | 0.1% | 0.1% | 0.0% | 0.0% | 0.0% | 0.0% | 0.0% | 0.2% |
| $A^2$ | Phones | 3.0% | 12.5% | 3.0% | 2.7% | 0.2% | 0.0% | 0.0% | 1.4% | 0.1% | 1.3% |
| | Conn. Cars | 16.5% | 23.8% | 0.9% | 0.0% | 0.2% | 0.1% | 0.0% | 0.0% | 0.0% | 0.1% |
| | Tablets | 0.5% | 6.5% | 0.8% | 0.2% | 0.0% | 0.0% | 0.0% | 0.0% | 0.1% | 1.0% |

As mentioned in §2, four control events (ATCH, DTCH, SRV_REQ, and S1_CONN_REL) can trigger a UE to transition among EMM/ECM states (REGISTERED, DEREGISTERED, CONNECTED, and IDLE), which cannot be captured by modeling each event type separately. Thus, we also model the sojourn time in those states for individual UEs using traditional probability distributions. Specifically, for each UE, within each interval, we replay the traffic trace while following the EMM and ECM state machines specified by 3GPP (Fig. 1 and Fig. 2), to calculate the duration of staying in each of the four UE states. For every UE-cluster/hour/device-type combination, we group the sojourn time across different UEs and different days, and use MLE to fit the grouped sojourn time.

**4.1.2 Validating traditional distribution-based modeling.** We first examine whether Poisson processes can model the inter-arrival time or the sojourn time for individual UE traffic in *each 1-hour interval per device type per UE-cluster*. Specifically, we apply two standard statistical tests, the Kolmogorov-Smirnov (K–S) test [32] and the Anderson-Darling ($A^2$) test [40]. The K–S test compares the maximum distance between the empirical cumulative distribution function (CDF) of the sample data and the theoretical CDF of the reference distribution (e.g., exponential distribution) [12]. It outputs the *K–S test statistic* (i.e., the supremum of the set of the distances) along with the corresponding *p-value*. A *p-value* of 0.05 or lower is considered statistically significant between the empirical distribution of the sample data and the reference distribution [45]. The $A^2$ test is a modification of the K–S test and gives more weight to the tails of the observation data [41]. It outputs the $A^2$ *test statistic*, the *critical values*, and the corresponding *significance levels* calculated for the reference distribution. The $A^2$ *test statistic* is used to compare against the *critical values* calculated specifically for the reference distribution to determine whether to accept or reject the null hypothesis under some *significance level*. In this paper, we focus on the *significant level* of 5%.

To examine fitting with other distributions (i.e., Pareto, Weibull and Tcplib), we repeat the same preprocessing procedure and only perform the K-S test to decide if the inter-event arrival time of each type of events per UE or the duration staying in each of the four UE states is drawn from one of those distributions. We skip the $A^2$ test, because it can only test against some common distributions at the moment (e.g., normal and exponential).

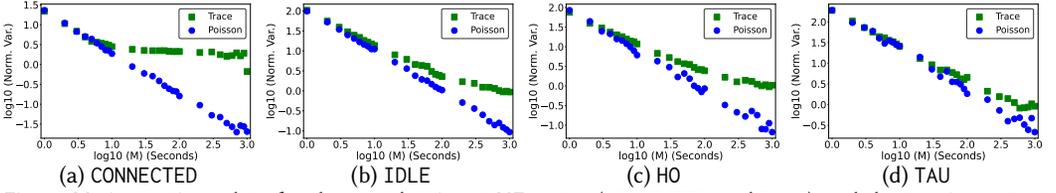

Fig. 7. Variance-time plots for the two dominant UE states (CONNECTED and IDLE) and the two important control events (HO and TAU) for one randomly-sampled UE cluster of phones.

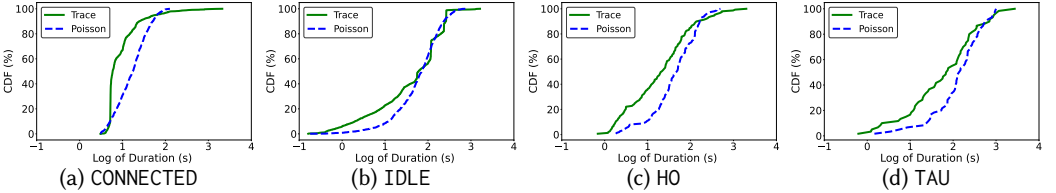

Fig. 8. Comparison of CDFs between real and fitted data (Poisson) for the CONNECTED and IDLE states and the HO and TAU events for one randomly-sampled UE cluster of phones.

**Results.** Table 3 shows that surprisingly the inter-arrival time of all six types of events (ATCH, DTCH, SRV_REQ, S1_CONN_REL, HO, and TAU) cannot be modeled as Poisson processes for each UE cluster of all three types of devices, since below 5.0% of the 1-hour intervals pass the K–S test and below 23.8% for the $A^2$ test for the exponential distribution. We note that there are limit theorems in the literature [14] stating that the superposition of many independent point processes approaches a Poisson process. However, such results are for the inter-arrival time of the *aggregate* arrival process, and thus do not imply that the *individual* point process can be modeled as a Poisson process. Since, in this paper, we are studying the inter-arrival time for each individual UE (i.e., individual point process), our findings do not contradict with [14].

Table 3 also shows that below 1.4% of the intervals pass the two statistical tests for the duration of UE staying in the four EMM/ECM states (REGISTERED, DEREGISTERED, CONNECTED and IDLE), for all three types of devices.

In summary, our analysis shows that *the inter-arrival time of the control-plane events, as well as the sojourn time in the UE states for the cellular network, cannot be modeled as Poisson processes*. We also find that they cannot be modeled by other traditional probability distributions (i.e., Pareto, Weibull and TCPlib). The details can be found in Appendix B.1.

### 4.2 Why do Traditional Probability Distributions Fail to Model Individual UE Traffic?

To understand why traditional probability distributions fail to model either the sojourn time in the four UE states or the inter-arrival time for different control events, we zoom into the two dominant UE states, CONNECTED and IDLE, and the two important types of control events, HO and TAU, and perform in-depth analysis: (1) How well the Poisson distribution can model an essential property of the control-plane traffic, its burstiness; (2) Whether the upper and lower tails of the observed inter-arrival time distribution can be captured by the Poisson distribution.

**Burstiness.** We first calculate the variance of numbers of events over different time scales, known as *variance-time plot* [19, 26], per UE-cluster, per hour, per device-type, to assess how well the Poisson distribution can model the burstiness of control-plane traffic. Specifically, we first divide the timeline into 100ms intervals. We count the number of events per 100ms interval for every type of device and event. Next, we consider different time scales $M$, ranging from 1 to $10^3$ seconds. For each time scale of $M$ seconds, we calculate the average number of events per 100ms for every $M$-second window $i$, denoted as $k_i$, and then calculate the mean and variance of this metric across all $M$-second windows, denoted as $\overline{k_i}$ and $\hat{k_i}$, respectively. We then normalize $\hat{k_i}$ by the square of $\overline{k_i}$ for the observation and the reference distribution, respectively.

Figure 7 shows that for one randomly-sampled UE clusters of phones,[5] the sojourn time in CONNECTED and IDLE exhibits stronger burstiness than the fitted Poisson models across the time scale ranging from 10 to $10^3$ seconds. The differences in the log-scale normalized variance between the real-world trace and the fitted exponential distribution are 0.43∼2.00 and 0.18∼1.00 for CONNECTED and IDLE, respectively. For HO and TAU, although they happen much less frequently than SRV_REQ and S1_CONN_REL, which trigger the state change between CONNECTED and IDLE, their inter-arrival time still has stronger burstiness than the fitted Poisson models across the time scale from 10 to $10^3$ seconds. The differences in the log-scale normalized variance between the real-world trace and the fitted exponential distribution are 0.20∼1.20 and -0.04∼0.63 for HO and TAU, respectively.

**Inter-arrival tails.** A more direct way of understanding why a traditional probability distribution cannot model a traffic trace is to examine whether the entire range of the observed inter-arrival time in the trace can be captured by the Poisson model [37]. We compare the CDF of the observed data with that of the fitted Poisson (exponential) distribution—over the same 1-hour interval for the sojourn time in CONNECTED and IDLE and the inter-arrival time of HO and TAU.

Figure 8 shows the results for the same two randomly sampled UE clusters of phones as above. We see that the exponential distribution fails to adequately capture the entire range of the four quantities observed in the trace. In particular, for CONNECTED, the maximum sojourn time is around 2106.94 seconds, much higher than that of the fitted exponential distribution, i.e., 156.35 seconds. For IDLE, the minimum sojourn time is around 0.16 seconds, smaller than that of the fitted exponential distribution, i.e., 0.40 seconds. For HO, the inter-arrival duration ranges from 0.69 to 1988.18 seconds, while the fitted inter-arrival duration only ranges from 0.78 to 559.56 seconds. As with HO, TAU has inter-arrival time ranging from 0.62 to 2721.36 seconds, while the fitted inter-arrival time only varies from 2.71 to 723.26 seconds.

### 4.3 How to Model Individual UE Traffic, the Right Way?

To capture the key characteristics of the mobile network control-plane traffic discussed above, we propose *a two-level state-machine-based traffic model* for each UE cluster (derived from an adaptive clustering scheme) that is also dependent on the hour-of-the-day and the device type. Instantiating such a state-machine-based model per UE-cluster, per hour, and per device-type allows us to capture both the dependence among events and the traffic diversity in these three dimensions.

In the following subsection, we first describe how to capture the dependence among different events using a two-level state machine. We then elaborate on how we derive key parameters of the state-machine-based model, i.e., the duration of a UE staying in each state before switching to the next state and the probability of transitioning from one state to the next state. We finally discuss how to model the first event and its start time for every UE, when synthesizing a new trace.

#### 4.3.1 Capturing the dependence among events in each UE. 
We observe that the six types of control events actually fall into two categories that have different dependence on each other; those that trigger a UE to transition among different UE states (denoted as Category-1 events), and those that do not but have complex dependence on UE states (denoted as Category-2 events).

Category-1 events include ATCH, DTCH, SRV_REQ, and S1_CONN_REL, which cause the UE state to switch from one state to another following the EMM and ECM state machines. We observe that when a UE changes from DEREGISTERED to REGISTERED, it always enters CONNECTED at the same time. Therefore, the EMM and ECM state machines can be merged as one state machine that captures all Category-1 events, as shown in Figure 3. We denote the merged state machine as the *EMM–ECM state machine*.

---
[5]The results for other UE clusters and device types are similar and omitted due to page limit.

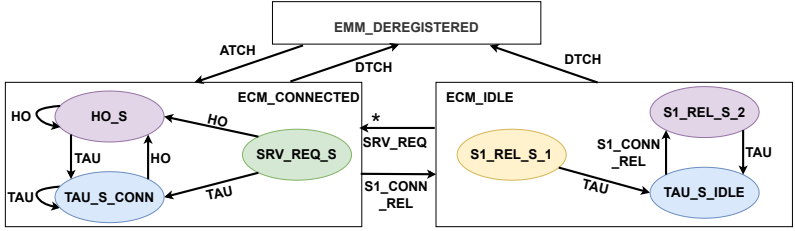

Fig. 9. Proposed two-level hierarchical state machine. The rectangles represent states at the top level. The ovals represent states at the bottom level. The arrow with the star denotes the transition SRV_REQ can only start from S1_REL_S_1 and S1_REL_S_2, while the opposite arrow denotes the transition S1_CONN_REL can start from any of the states in CONNECTED.

In contrast with Category-1 events, Category-2 events (HO and TAU) do not change the UE state. However, these events still have intricate dependence on other control events as follows. (1) For HO, it only happens after the UE enters CONNECTED triggered by SRV_REQ. (2) For TAU, although it can happen in both CONNECTED and IDLE, it sometimes follows HO in CONNECTED. This is because when the UE moves and switches from one cell to another cell, it can also enter another tracking area (a new set of cells). However, TAU does not necessarily follow HO in CONNECTED, for example, when the UE reselects to LTE from 3G or the UE re-registers to LTE after the fallback for circuit-switched services, etc. (3) In addition, in the IDLE state, after TAU, S1_CONN_REL always happens in order to release the signaling resources assigned for the last TAU. None of these dependencies can be easily captured by the EMM–ECM state machine. The key challenge here is that these events can happen in either CONNECTED or IDLE or both, and with different dependencies on other events.

To overcome the above challenges to capture the dependence for Category-2 events under the CONNECTED and IDLE states, we introduce two fine-grained sub-state machines that are embedded inside the CONNECTED and IDLE states of the EMM–ECM state machine, which effectively refines the EMM–ECM state machine into a two-level hierarchical state machine. In each sub-state machine, each state corresponds to the event that happens right before entering this state, e.g., the SRV_REQ_S state is entered after a SRV_REQ event. Each edge corresponds to a Category-2 control event, just like each edge in the EMM–ECM state machine corresponds to a Category-1 control event. Specifically, we define six new states in the sub-state machines, including HO_S, TAU_S_CONN, TAU_S_IDLE, SRV_REQ_S, S1_REL_S_1, and S1_REL_S_2. TAU_S_CONN and TAU_S_IDLE are needed to distinguish the state entered after a TAU event in the CONNECTED and IDLE states of the EMM–ECM state machine, respectively. Unlike S1_REL_S_1, which represents the UE switching from CONNECTED to IDLE, S1_REL_S_2 is needed to capture the unique behavior of TAU in the IDLE state.

Figure 9 (bottom left) shows the sub-state machine inside CONNECTED. After SRV_REQ happens and changes the UE to the SRV_REQ_S state, the UE can either trigger HO to enter HO_S or trigger TAU to enter TAU_S_CONN. When the UE is in HO_S, there are two possible transitions. If the next event is HO, it causes the UE to self-loop back to HO_S. If the next event is TAU, it changes the UE to TAU_S_CONN. Similarly, when the UE is in the TAU_S_CONN state, the next event is either TAU, which self-loops the UE to TAU_S_CONN, or HO, which changes the UE to HO_S.

Figure 9 (bottom right) shows the other sub-state machine inside IDLE. After S1_CONN_REL happens and changes the UE to the S1_REL_S_1 state, the UE can trigger TAU and then enters TAU_S_IDLE. Then, S1_CONN_REL is triggered to make the UE enter the S1_REL_S_2 state. When another TAU happens, UE changes back to the TAU_S_IDLE state.

In essence, those two sub-state machines are the second-level refined state machines embedded in the EMM–ECM state machine. Further, they can run concurrently with the top-level EMM–ECM state machine. For example, suppose that the UE is moving. The UE may switch from CONNECTED to IDLE, if the UE does not transmit data for some seconds. However, in the meantime, regardless of whether the UE is in CONNECTED or IDLE, TAU can happen, as shown in Figure 9.

### 4.3.2 Modeling the duration of a UE staying in the current state and the probability of transitioning to the next state.
To instantiate the state machine model, i.e., to derive the parameters of the model, we adopt the *Semi-Markov model* [47], which is a multi-state model for a continuous time stochastic process with state transitions. In contrast with the Markov model, it does not assume that the sojourn time in the same state follows the exponential distribution with a constant hazard rate, which, as shown in §4.1, is not applicable to mobile network control traffic. The Semi-Markov model generally models (1) the probability of the state transition from state $x$ to state $y$: $p_{xy} = \mathbb{P}(S_{i+1} = y \mid S_i = x)$, where $S_i$ and $S_{i+1}$ represent the states of two consecutive steps $i$ and $i+1$, respectively; (2) the duration of staying in state $x$ before switching to state $y$, as a random variable with the CDF: $F_{xy}(t) = \mathbb{P}(T_{i+1} - T_i \leq t \mid S_i = x, \ S_{i+1} = y)$, where $T_i$ and $T_{i+1}$ represent the time of the process switching to the state $S_i$ and $S_{i+1}$, respectively.

To model the sojourn time in one state before switching to another state, we first collect the sojourn time for the same transition across all UEs. We then derive one CDF for the sojourn time of the same transition, since traditional probability distributions fail to model the duration, not only for the EMM and ECM states (§4.1), but also for the new states in the two-level state machine (Appendix B.2). To model the transition probability from one state to another, we count numbers of transitions across all UEs and then calculate $p_{xy}$ for every transition (i.e., each edge in the state machine). If there is only one outbound edge, $p_{xy}$ of the corresponding state transition is 1.

### 4.3.3 Capturing the traffic diversity in the two-level state machine model.
As shown in §3, the UEs of the same device type have diverse traffic for each type of control-plane events and the distributions of the UEs are highly skewed for all three types of devices. As a result, using a single model for all UEs fails to reproduce such diversity, as the sojourn time and the transition probabilities of more active UEs dominate corresponding CDFs.

To capture traffic diversity, the obvious choice is to *cluster the traffic for each event type* of the same device type, and then derive a model per cluster for every event type for every device type. However, such an approach does not work, because it ignores the *dependence among different types of events* and thus the *dependence among models of different event types*. For example, the two dominant events (SRV_REQ and S1_CONN_REL) are dependent on each other, following the ECM state machine. They should have an identical number of occurrences during a given time period for any UE. Separately clustering different event types cannot directly capture such dependence.

To capture the diversity of $p_{xy}$ and $F_{xy}(t)$ across similar UEs (e.g., of the same device type, with the same traffic intensity) along with the dependence among events of the same UE, we propose to *cluster the UEs based on their traffic similarity*, and derive different models for different event types per UE-cluster for every device type. Modeling per UE-cluster inherently captures the dependence among different types of events.

Ideally, each cluster should cover as many similar UEs as possible. However, it is challenging to achieve both a large number of UEs and high similarity within the same cluster, because of the highly diverse and skewed control-plane traffic over all UEs for all three types of devices.

To strike a balance between those two goals, we propose *an adaptive clustering scheme* to recursively segregate the UEs into different clusters, until the UEs in the cluster have roughly similar traffic patterns or the number of UEs in the cluster is small enough. To quantify the similarity of UE traffic, we focus on two dominant events, SRV_REQ and S1_CONN_REL, which contribute to 84.1%~93.0% of the total control events for all three types of devices. We extract two features for each event type to characterize the UE traffic: (1) number of control events; (2) standard deviation of the duration staying in the state (CONNECTED or IDLE).

The recursive adaptive scheme works as follows. It is first invoked for the complete feature space and all UEs are grouped into one cluster. At each invocation, it checks for one cluster if the features of UEs are similar (i.e., the difference between maximum and minimum value should be smaller than some threshold $\theta_f$ for every feature), or if the number of UEs is smaller than another threshold $\theta_n$. If neither one is satisfied, the procedure cuts the current feature space into 4 equal-sized sub-feature spaces, and the UEs that fall into the same sub-feature space are grouped into a sub-cluster. Effectively, the recursively partitioned sub-feature space forms a quadtree, and the UEs in the sub-feature space that is not partitioned further are grouped into a final cluster used to model the sojourn time and the transition probabilities for different state transitions. We experimentally find that a $\theta_f$ value of 5 for all features and a $\theta_n$ value of 1000 are sufficient to segregate the UEs in the input trace into groups of UEs with sufficiently dissimilar behavior.

Using the above thresholds, we generated 574, 199 and 70 UE clusters per hour on average for phones, connected cars, and tablets, respectively. As a result, we instantiated a total of 20,216 two-level state machines, one for each combination of UE cluster, hour-of-day and device type.

### 4.3.4 Modeling the first event and corresponding start time.
In addition to modeling the sojourn time of the UE staying in one state and the probability of transitioning to the next state, the traffic generator also needs to decide on the very first event and corresponding start time for every UE, every time it synthesizes a new trace, e.g., given a starting hour of the day.

To do this, for each hour of the input trace, we collect the first event and the start time of all the UEs per UE-cluster discussed above, and derive the probabilities of different event types as the first event for the hour as well as the distribution of the start time within the hour.

### 4.3.5 Putting it all together.
We proposed a two-level state-machine-based control traffic model for mobile networks as shown in Figure 9. The top-level model captures the behavior of Category-1 control events (ATCH, DTCH, SRV_REQ, and S1_CONN_REL). The two bottom-level models are for Category-2 events (HO and TAU). To capture the diversity in control traffic, we proposed to instantiate the model, i.e., of its parameters, once for each type of device for every hour and for each cluster of UEs that exhibit similar traffic behavior. In particular, for each type of device, in each hour, we first apply the adaptive clustering scheme to group the UEs in the input trace into different clusters. Then, for each UE cluster, we leverage the Semi-Markov model to instantiate a two-level model, i.e., annotating each state with the duration of a UE staying in that state and the probabilities of the transitions from that state to other states. Finally, we derive the first-event model for each combination of (UE cluster, hour, device type), which models the first event types and corresponding start time within each hour needed to bootstrap the traffic generator, and we model the distribution of the UEs over different clusters, which is needed to determine the fractions of UEs that should use the state machine model for each (UE cluster, hour, device type) in the overall traffic generator.

## 4.4 How to Use the Proposed Traffic Models to Generate New Traces?

The traffic generator uses the two-level state-machine-based traffic model per device type, per hour, per UE-cluster to generate new control traffic traces for any given number of UEs, starting at any given hour. To synthesize a new trace for $K$ UEs starting at hour $H$, the main traffic generator runs $K$ instances of the per-UE traffic generator concurrently. Each per-UE traffic generator uses the traffic models (for different hours) of a given UE cluster, according to the distribution of the UEs in the modeled trace, e.g., if 33% of the UEs belong to Cluster X, then 33% of the per-UE traffic generators will be running the state machine for Cluster X. The per-UE traffic generator first decides the first event and the corresponding start time within hour $H$ by following the first-event model, and then starts driving the per-hour state machine for that UE cluster for that device type one hour after another hour.

Table 4. Breakdown comparison of events between the real traces and the synthesized traces using the baseline and our method under the two validation scenarios.

| | Scenario 1 (38,000 UEs) | | | | | | | | | Scenario 2 (380,000 UEs) | | | | | | | | |
|---|---|---|---|---|---|---|---|---|---|---|---|---|---|---|---|---|---|---|
| | Phones | | | Conn. Cars | | | Tablets | | | Phones | | | Conn. Cars | | | Tablets | | |
| | Real | Base | Ours | Real | Base | Ours | Real | Base | Ours | Real | Base | Ours | Real | Base | Ours | Real | Base | Ours |
| ATCH | 0.1% | 0.1% | 0.1% | 0.8% | 0.21% | 1.2% | 0.5% | 0.4% | 1.0% | 0.1% | 0.2% | 0.1% | 0.8% | 0.2% | 1.1% | 0.5% | 0.3% | 1.0% |
| DTCH | 2.0% | 0.3% | 2.1% | 6.1% | 0.42% | 7.1% | 2.3% | 0.7% | 3.1% | 2.0% | 0.4% | 2.1% | 6.2% | 0.4% | 6.8% | 2.3% | 0.6% | 3.0% |
| SRV_REQ | 45.6% | 9.3% | 46.9% | 37.6% | 4.61% | 42.7% | 46.1% | 17.7% | 46.1% | 45.5% | 10.0% | 46.9% | 38.2% | 4.4% | 42.7% | 46.3% | 14.2% | 46.3% |
| S1_CONN_REL | 46.8% | 9.6% | 47.8% | 42.4% | 4.99% | 44.5% | 47.8% | 18.3% | 47.5% | 46.8% | 10.4% | 47.9% | 42.2% | 4.8% | 44.6% | 47.7% | 14.7% | 47.5% |
| HO (CONN.) | 3.5% | 24.9% | 1.9% | 7.1% | 17.80% | 2.5% | 1.6% | 14.7% | 1.3% | 3.5% | 23.4% | 1.8% | 7.5% | 20.3% | 2.7% | 1.9% | 12.0% | 1.2% |
| HO (IDLE) | 0.0% | 32.6% | 0.0% | 0.0% | 38.27% | 0.0% | 0.0% | 29.5% | 0.0% | 0.0% | 35.0% | 0.0% | 0.0% | 38.5% | 0.0% | 0.0% | 21.7% | 0.0% |
| TAU (CONN.) | 0.7% | 11.5% | 0.3% | 1.3% | 17.14% | 0.5% | 0.3% | 6.7% | 0.2% | 0.7% | 11.3% | 0.3% | 1.3% | 16.4% | 0.5% | 0.3% | 5.3% | 0.2% |
| TAU (IDLE) | 1.2% | 11.8% | 0.8% | 4.7% | 16.56% | 1.5% | 1.5% | 12.0% | 0.8% | 1.5% | 9.3% | 0.8% | 3.8% | 15.1% | 1.6% | 1.1% | 31.2% | 0.7% |

In more detail, each per-UE traffic generator first samples the first event and the corresponding start time from the first-event model. It then runs the per-hour two-level state machine one after another, starting from hour $H$. For each level of the two-level state machine, it keeps a timer to track the time the UE stays in the current state. Whenever the UE enters a state $x$, the traffic generator (1) follows the probabilities $\{p_{xy} \mid y \in$ all states coming from $x\}$ to decide the next state $y$ and the corresponding event $e$ that triggers the state transition from $x$ to $y$; and (2) follows the CDF ($F_{xy}$) to decide the duration $D$ for staying in the current state $x$. The traffic generator sets the timer for $D$ seconds and starts it. When the timer expires, the UE generates event $e$ and enters state $y$, and the traffic generator repeats the process above. In addition, when the state at the top level is changed, then for the bottom level, the traffic generator (1) drops its next event, which was decided to happen later in the past top-level state, (2) resets its timer and (3) starts running the sub-state machine corresponding to the new state of the top level, e.g., Figure 9 (bottom left) if the top-level state machine enters CONNECTED.

### 4.5 How to Validate the Proposed Traffic Models (and Traces)?

In this subsection, we validate the proposed two-level state machine-based traffic model by comparing it with a baseline method that uses the fitted Poisson distributions. To fairly compare the performance of those two methods, we apply to the baseline the same UE clusters as the two-level state machine-based model for all three device types. For each UE cluster, the baseline method considers two groups of event types differently: (1) ATCH, DTCH, SRV_REQ, and S1_CONN_REL; (2) HO and TAU. For the first group of event types, the baseline simply follows the EMM−ECM state machine (Fig. 3), and uses the Poisson distribution to model the sojourn time of the UE staying in DEREGSITERED, CONNECTED and IDLE. As for the second group, the baseline also uses the Poisson distribution to model the inter-arrival time of HO and TAU, separately. Meanwhile, the baseline method models the first event and the corresponding start time for the two groups of event types, respectively. To synthesize a new trace for $K$ UEs, the baseline method runs $K$ instances for each UE. Each per-UE traffic generator runs two threads, one for each group of event types in parallel.

To assess the scalability of the proposed method and the baseline method, we consider two different validation scenarios with different numbers of UEs: 38,000 UEs and 380,000 UEs, i.e., about 1× and 10× more than the UEs that we used to estimate the model parameters. Using either method, we synthesize two traces for those two scenarios for one of the busy hours on a randomly-chosen day in August 2022. We then compare the synthesized traces with the real traces that we randomly sampled for the corresponding numbers of UEs and hour of the day.

To evaluate the fidelity of the synthesized traces using the two methods, we consider metrics from two perspectives: *macroscopic* and *microscopic*. From the macroscopic perspective, we compare the breakdown of the synthesized trace into different control-plane events with that of the real trace for each type of devices. From the microscopic perspective, we zoom into the events generated for each UE, and compare per-UE traffic behavior in two ways: (1) numbers of events per UE for different state transitions; (2) sojourn time of each UE staying in one state before transiting to another state for different transitions.

**4.5.1 Macroscopic analysis.** We first compute the breakdown of the synthesized events and compare it with that of the real events for each type of devices under Scenario 1 with 38K UEs. Table 4 (left) shows that the proposed method synthesizes a trace whose breakdowns of control-plane events for all three types of devices are much closer to those of the real trace, than those of the trace generated by the baseline. For SRV_REQ and S1_CONN_REL, the percentages in the synthesized trace by our method differ only by 1.1%~1.3% (phones), 2.1%~5.0% (connected cars) and 0.1%~0.3% (tablets) from that in the real trace, while their percentages in the baseline trace differ by 36.4%~37.2% (phones), 33.0%~37.5% (connected cars) and 28.4%~29.5% (tablets).

Table 4 also shows that for HO events, which should happen only in the CONNECTED state, our method reproduces similar fractions of HO in CONNECTED and stops generating HO in IDLE. Specifically, in CONNECTED, the absolute differences in the percentages between our synthesized trace and the real trace are within 1.7% (phones), 4.6% (connected cars) and 0.3% (tablets), much smaller than those between the traces synthesized by the baseline and the real traces, i.e., 21.3% (phones), 10.7% (connected cars) and 13.2% (tablets). In IDLE, the baseline mistakenly generates 35.0%, 38.5% and 21.7% of the total events as HO, because the state dependence for HO is not captured by the baseline. For TAU, which can happen in both CONNECTED and IDLE, our method can synthesize TAU in different ECM states correctly. The absolute differences in the percentages between the synthesized and real traces for TAU in CONNECTED and IDLE are 0.3%~0.5%, 0.8%~3.1% and 0.1%~0.7% for phones, connected cars, and tablets, respectively. In contrast, the baseline-synthesized traces have much larger differences from the real trace, i.e., 10.5%~10.8%, 11.9%~15.9% and 6.4%~10.6% for the three types of devices, respectively. The results above show that our proposed two-level state machine-based traffic model using the Semi-Markov model can capture the state dependence well.

We next compare the breakdown of the synthesized events and the real events for the three types of devices under Scenario 2 with 380K UEs, about 10× more UEs than what we used for modeling, as shown in Table 4 (right). We find that, compared with the real trace, the traffic generator using our proposed traffic model synthesizes a trace with very similar breakdowns of control-plane events for all three types of devices. The percentages of SRV_REQ and S1_CONN_REL in our synthesized trace are within 1.0%~1.4%, 2.5%~4.5% and 0.0%~0.1% of that in the real trace, while those in the trace generated by the baseline are 35.4%~36.5%, 33.9%~37.4% and 32.1%~33.0% larger than those in the real trace. As with Scenario 1, the trace synthesized by our proposed traffic model exhibits up to 1.7%, 4.9% and 0.7% differences for HO, and up to 0.3%~0.6%, 0.8%~2.2% and 0.1%~0.4% differences for TAU in CONNECTED and IDLE, compared with the real trace. In contrast, the percentages of HO and TAU in CONNECTED and IDLE generated by the baseline differ by 7.9%~35.0%, 11.3%~38.5% and 5.0%~30.1% from those in the real trace. These results suggest that our traffic model can synthesize realistic traces much larger than the trace used to seed the model, e.g., for a 10× larger UE population.

**4.5.2 Microscopic analysis.** We next evaluate the microscopic fidelity of the synthesized traces in two ways: (1) numbers of events per UE for different event types; (2) sojourn time of each UE staying in one state before transiting to another state for different state transitions. We compare the distributions between real and synthesized traces by first deriving the CDFs for both traces and then computing the maximum y-distance, i.e., the distance along the y-axis (the probability of any "event" less than the x-axis value) of the two CDFs, which is a conservative way to measure the fidelity of the synthesized trace.

Table 5. Maximum y-distance between CDFs of the synthesized and real traces for the numbers of SRV_REQ/ S1_CONN_REL per UE and the sojourn time in CONNECTED/IDLE per UE under the two validation scenarios.

|  | Scenario 1 (38,000 UEs) | | | | | | Scenario 2 (380,000 UEs) | | | | | |
|---|---|---|---|---|---|---|---|---|---|---|---|---|
|  | Phones | | Conn. Cars | | Tablets | | Phones | | Conn. Cars | | Tablets | |
|  | Base | Ours | Base | Ours | Base | Ours | Base | Ours | Base | Ours | Base | Ours |
| SRV_REQ | 53.1% | 6.9% | 38.2% | 33.2% | 52.8% | 16.7% | 52.8% | 6.7% | 37.5% | 32.3% | 52.5% | 16.0% |
| S1_CONN_REL | 52.4% | 7.0% | 38.8% | 32.9% | 52.6% | 17.2% | 52.1% | 6.8% | 37.9% | 32.0% | 52.3% | 17.0% |
| CONNECTED | 30.2% | 6.3% | 25.0% | 9.4% | 23.4% | 2.7% | 31.0% | 6.1% | 23.5% | 6.5% | 23.1% | 2.1% |
| IDLE | 15.5% | 4.8% | 14.4% | 11.7% | 23.0% | 8.2% | 15.2% | 4.3% | 13.7% | 10.4% | 21.7% | 6.8% |

Table 6. Maximum y-distance between CDFs of the synthesized and real traces for the numbers of events per UE for the two groups of UEs per device type (inactive UEs / active UEs).

|  | Scenario 1 (38,000 UEs) | | | | Scenario 2 (380,000 UEs) | | | |
|---|---|---|---|---|---|---|---|---|
|  | Conn. Cars | | Tablets | | Conn. Cars | | Tablets | |
|  | Base | Ours | Base | Ours | Base | Ours | Base | Ours |
| SRV_REQ | 92.0%/46.4% | 24.7%/12.2% | 43.6%/23.6% | 20.7%/9.8% | 92.3%/47.2% | 25.3%/11.1% | 42.0%/22.0% | 22.7%/7.8% |
| S1_CONN_REL | 94.5%/53.6% | 23.1%/11.8% | 48.9%/26.0% | 28.4%/9.9% | 94.9%/54.3% | 22.8%/10.6% | 47.1%/26.1% | 30.8%/7.6% |

Since SRV_REQ and S1_CONN_REL, which switch the UE state between CONNECTED and IDLE, dominate the control events, i.e., accounting for 84.1%~93.0% of the total events shown in §3, we focus on these events and compare their distributions between the synthesized and real traces.

**Number of events per UE.** Table 5 shows that for phones, our traffic generator can synthesize similar numbers of SRV_REQ and S1_CONN_REL per UE under both validation scenarios; the maximum y-distance between CDFs of the synthesized and real traces is 6.7%~7.0% using our proposed traffic model, while it is 52.1%~53.1% using the baseline. However, for connected cars and tablets, SRV_REQ and S1_CONN_REL synthesized by our proposed model have the maximum y-distance of 32.0%~33.2% (connected cars) and 16.0%~17.2% (tablets). In contrast, the CDFs for SRV_REQ and S1_CONN_REL generated by the baseline have even larger maximum y-distance of 37.5%~38.8% (connected cars) and 52.3%~52.8% (tablets) from those in the real trace.

To understand the high maximum y-distance for connected cars and tablets, we zoom into the CDFs of the number of SRV_REQ/S1_CONN_REL per UE in the real and synthesized traces for all three types of devices (shown in Appendix C). We find that for both SRV_REQ and S1_CONN_REL in the synthesized trace by our proposed traffic model, the high maximum y-distance of connected cars and tablets is caused by the UEs that generate only 1 event occurrence during the selected hour, while the traffic generator predicts 2 occurrences. To quantitatively verify it, for each type of devices, we split the UEs into two groups: (1) inactive UEs with fewer than or equal to 2 occurrences; (2) active UEs with more than 2 occurrences during the selected hour. We calculate the maximum y-distance between the CDFs of the synthesized and real traces for the two groups of UEs separately, for each device type. Table 6 shows that for synthesized SRV_REQ and S1_CONN_REL using our proposed model, the maximum y-distance for active UEs with more than 2 events per UE is 10.6%~12.2% (connected cars) and 7.6%~9.9% (tablets) under the two validation scenarios, while the maximum y-distance for inactive UEs with fewer than or equal to 2 events per UE is 22.8%~25.3% (connected cars) and 20.7%~30.8% (tablets) under the two validation scenarios. However, for the trace generated by the baseline, the maximum y-distance for active UEs with more than 2 events per UE is 46.4%~54.3% (connected cars) and 22.0%~26.0% (tablets) under the two validation scenarios, while the maximum y-distance for inactive UEs with fewer than or equal to 2 events per UE is 92.0%~94.9% (connected cars) and 42.0%~48.9% (tablets) under the two validation scenarios. The results suggest that our proposed traffic model only mis-predicts the number of events by 1 during the selected hour for connected cars and tablets, much better than the baseline model. We argue that such a difference by just 1 event per UE within a single hour for inactive connected cars and tablets is acceptable.

Table 7. Mapping of primary control-plane event types between 4G (left) and 5G (right).

| 4G | 5G |
| --- | --- |
| ATCH | REGISTER (Registration) |
| DTCH | DEREGISTER (Deregistration) |
| SRV_REQ | SRV_REQ (Service Request) |
| S1_CONN_REL | AN_REL (AN Release) |
| HO | HO (Handover) |
| TAU | – |

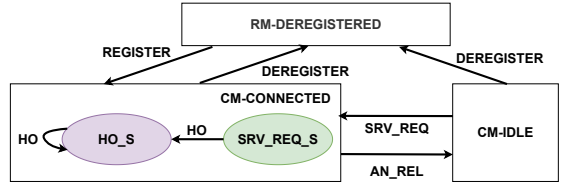

Fig. 10. The adjusted two-level state machine for 5G.

**Duration of staying in one state before switching to the next state per UE.** Table 5 also shows that our proposed method can simulate more accurate distributions of the sojourn time than the baseline method for the two dominant UE states (CONNECTED and IDLE); the maximum y-distance for our method for both states is similar under both validation scenarios, i.e., 4.3%~6.3% (phones), 6.5%~11.7% (connected cars) and 2.1%~8.2% (tablets), while it is 15.2%~31.0% (phones), 13.7%~25.0% (connected cars) and 21.7%~23.4% (tablets) for the baseline.

## 5 TOWARDS MODELING CONTROL-PLANE TRAFFIC FOR 5G NETWORKS

As mentioned in §4, one of the important goals of modeling control-plane traffic is to generate control-plane traffic not only experienced by today's MCN (e.g., 4G) but also for NextG, tomorrow. In this section, we present a methodology to adjust our proposed traffic model from LTE to 5G.

We first carefully studied the 3GPP specifications of 4G [3] and 5G [4, 5]. We find that there exists a one-to-one mapping of event types between 4G and 5G for all primary types of control events, except TAU as summarized in Table 7. As for the four UE states in LTE (REGISTERED, DEREGISTERED, CONNECTED, and IDLE), there are also four corresponding UE states in 5G (RM-REGISTERED, RM-DEREGISTERED, CM-CONNECTED, and CM-IDLE). Such high similarity of control-plane event types and UE states between 4G and 5G enables us to easily adjust our proposed two-level state-machine-based model from 4G to 5G as follows.

To adjust the two-level state machine, we simply remove the states and related transitions of TAU from the state machine for 4G (Fig. 9) to form a new two-level state machine for 5G (Fig. 10).

There are two ways to instantiate the adjusted two-level state machine model for 5G. (1) When a large-scale 5G control-plane trace 5G is available, we just need to follow the same methodology (§4.3) and derive the parameters for the final traffic model. (2) When such a 5G trace is not available, e.g., due to privacy concerns, we can scale the parameters of the derived traffic model for 4G in this work to derive the model parameters for the 5G model. Specifically, if we can estimate the scaling of each type of events when a UE switches from 4G to 5G, we can use the scaling factors to adjust the parameters in the 4G model for the 5G model. For example, recent measurement studies (e.g., [20]) have shown UEs tend to incur on average 4.6× more HO events when switching from 4G to 5G mmWave (NSA). Such frequency changes of some types of events can be used to recalculate the transition probabilities and the sojourn time in the two-level hierarchical state machine (Fig. 10).

## 6 CONCLUSION

In this paper, we carry out, to our knowledge, the first in-depth characterization of control-plane traffic, using an extensive real-world control-plane trace collected at an LTE MCN of a large mobile operator. We observe that control-plane events exhibit significant device-type and temporal diversity among UEs and the event-dependence in each UE. Building on the above characterization of the real LTE control-plane trace, we propose to model individual UE traffic, by developing a two-level hierarchical state-machine-based control-plane traffic model for each UE cluster derived from our adaptive clustering scheme based on the Semi-Markov Model. Our validation results show that our model outperforms traditional probability-based model and can synthesize traces for a much larger number of UEs than the trace we use to instantiate the model. We finally show how our model can be easily adjusted from LTE to 5G to support modeling 5G control-plane traffic. We will open-source the developed control-plane traffic models to the community to stimulate further research on MCN control-plane design and optimization for 4G/5G and beyond.

# APPENDICES

# A METHODOLOGY OF CHARACTERIZING THE TEMPORAL DIVERSITY AMONG UES

This appendix provides the methodology of how we generate the box plots (shown in Figure 4–6) to analyze the diversity of numbers of events per UE as well as average duration of each UE staying in different EMM/ECM states across different hours of the day (discussed in §3.2).

Table 8. Percentages of the 1-hour intervals whose inter-arrival time of different event types or sojourn time in the four EMM/ECM states passes the K–S test for other distributions (in addition to Poisson).

| Test | Device Type | ATCH | DTCH | SRV_REQ | S1_CONN_REL | HO | TAU | REG. | DEREG. | CONN. | IDLE |
|---|---|---|---|---|---|---|---|---|---|---|---|
| Pareto | Phones | 0.0% | 0.0% | 0.1% | 0.7% | 7.8% | 10.2% | 0.0% | 0.0% | 2.9% | 5.0% |
|  | Conn. Cars | 0.0% | 0.0% | 0.0% | 0.3% | 3.9% | 9.7% | 0.0% | 0.2% | 0.0% | 0.7% |
|  | Tablets | 0.5% | 0.0% | 0.5% | 1.2% | 5.0% | 6.0% | 0.0% | 0.0% | 3.4% | 2.7% |
| Weibull | Phones | 22.0% | 40.0% | 7.5% | 5.5% | 10.2% | 10.2% | 0.0% | 1.4% | 1.1% | 5.3% |
|  | Conn. Cars | 2.5% | 0.0% | 1.9% | 0.8% | 18.7% | 10.4% | 0.1% | 0.0% | 0.0% | 0.5% |
|  | Tablets | 6.6% | 20.7% | 3.3% | 3.3% | 11.8% | 7.3% | 0.0% | 3.4% | 0.9% | 1.7% |
| Tcplib | Phones | 0.0% | 0.0% | 0.4% | 0.4% | 0.3% | 0.3% | 0.0% | 0.0% | 0.3% | 0.3% |
|  | Conn. Cars | 0.0% | 0.0% | 0.0% | 0.0% | 1.5% | 1.5% | 0.0% | 0.0% | 0.0% | 0.0% |
|  | Tablets | 0.0% | 0.0% | 0.5% | 0.6% | 0.0% | 0.0% | 0.0% | 0.0% | 0.3% | 0.3% |

Table 9. Percentages of 1-hour intervals over all clusters that pass the two standard statistical tests for the state transitions (denoted as outbound state–trigger event) in the two second-level state machines.

| Test | Device Type | SRV_REQ_S-HO | HO_S-HO | TAU_S_C-HO | SRV_REQ_S-TAU | TAU_S_C-TAU | HO_S-TAU | S1_REL_1-TAU | S1_REL_2-TAU | TAU_S_I-S1_REL |
|---|---|---|---|---|---|---|---|---|---|---|
| Poisson (K–S) | Phones | 0.0% | 0.0% | 0.0% | 0.0% | 0.0% | 0.0% | 0.0% | 0.0% | 0.0% |
|  | Conn. Cars | 0.0% | 0.0% | 0.0% | 0.0% | 0.0% | 0.0% | 0.0% | 0.0% | 0.0% |
|  | Tablets | 0.0% | 0.0% | 0.0% | 0.0% | 0.0% | 0.0% | 0.0% | 0.0% | 0.0% |
| Poisson ($A^2$) | Phones | 0.0% | 0.2% | 0.9% | 0.0% | 0.0% | 0.0% | 0.0% | 0.0% | 0.0% |
|  | Conn. Cars | 0.0% | 0.1% | 0.0% | 0.0% | 0.0% | 2.9% | 0.0% | 0.0% | 0.0% |
|  | Tablets | 0.0% | 0.7% | 0.0% | 0.0% | 0.0% | 0.0% | 0.8% | 0.0% | 0.0% |
| Pareto (K–S) | Phones | 24.5% | 5.9% | 8.8% | 0.0% | 0.0% | 0.0% | 2.5% | 1.6% | 0.0% |
|  | Conn. Cars | 1.3% | 4.8% | 7.8% | 0.0% | 0.0% | 1.3% | 0.0% | 0.3% | 0.0% |
|  | Tablets | 14.2% | 12.8% | 10.0% | 0.0% | 0.0% | 1.6% | 0.0% | 0.6% | 0.0% |
| Weibull (K–S) | Phones | 6.5% | 5.2% | 4.3% | 0.0% | 0.0% | 0.4% | 21.5% | 12.3% | 0.0% |
|  | Conn. Cars | 18.1% | 15.6% | 15.0% | 0.0% | 0.0% | 17.0% | 1.6% | 0.9% | 0.0% |
|  | Tablets | 16.7% | 9.5% | 3.3% | 0.0% | 0.0% | 1.6% | 9.0% | 4.2% | 0.0% |
| Tcplib (K–S) | Phones | 0.5% | 0.2% | 1.2% | 0.0% | 0.0% | 0.0% | 0.0% | 0.0% | 0.0% |
|  | Conn. Cars | 0.0% | 0.4% | 0.0% | 0.0% | 0.0% | 0.0% | 0.3% | 0.0% | 0.0% |
|  | Tablets | 0.0% | 0.0% | 3.3% | 0.0% | 0.0% | 0.0% | 0.0% | 0.0% | 0.0% |

**Methodology.** To study the temporal diversity of number of events per UE, we first divide the entire trace into non-overlapping 1-hour intervals. Within each interval, we count number of events for every type of control events for each UE. We then group them in terms of the device type. We finally draw the box plot for the numbers of events per UE for all UEs of the same device type across different hours of the day for each type of events (shown in Figure 4). Thus, each box plot comprises 24 boxes, where each box is defined by the lower and upper quartiles of the data, the whiskers at the two ends represent it minimum and maximum, the red line in the center of the box is the median and the orange line is the average.

We follow the methodology above to plot for average duration of each UE staying in the EMM/ECM states (Figure 5) as well as percentages of TAU in CONNECTED per UE (Figure 6).

## B LIMITATIONS OF TRADITIONAL PROBABILITY DISTRIBUTIONS

In this appendix, we present additional results to show how traditional probability distributions fail to model the inter-arrival time between events, and the sojourn time in the four EMM/ECM states (Appendix B.1) as well as the new states in the proposed traffic model (Appendix B.2), for individual UE traffic.

## B.1 Can Other Traditional Probability Distributions Model Individual UE Traffic?

We first discuss the results showing the individual UE traffic (inter-arrival time of different types of events and the sojourn time staying in the four EMM/ECM states) cannot be fitted using the other traditional probability models, in addition to the Poisson model discussed in §4.1.

**Results.** Table 8 shows neither the inter-arrival time nor the sojourn time can be modeled by Pareto, Weibull, and Tcplib distributions for each UE cluster of all three types of devices. In particular, the Weibull distribution model achieves the largest percentage of the 1-hour intervals that pass the K-S test over all UE clusters, i.e., up to 40.0%, while the Pareto distribution and the Tcplib distribution have at most 10.2% and 1.5% of intervals that pass the K–S test.

## B.2 Can Traditional Probability Distributions Model the Sojoun Time in New States Proposed in Our Model?

To capture the state dependence among events, we propose a two-level state machine where the first-level state machine is the EMM–ECM state machine and the second-level state machine includes six new states and nine corresponding state transitions as shown in Fig. 9. We next discuss whether those new states can be modeled using traditional probability distributions.

Following the same methodology mentioned in §4.1 we first apply traditional probability distributions to the sojourn time of each state for each combination of UE clusters, 1-hour intervals, and device types. We then apply both the K–S test and the $A^2$ test to the Poisson distribution and apply only the K–S test to the other distributions, since the $A^2$ test can only be applied to some common distributions at the moment (e.g., normal and exponential).

**Results.** Table 9 shows all traditional distributions cannot properly model the sojourn time of UE staying in those states for each UE cluster of all three types of devices. For the Poisson distribution, close to zero intervals and up to 2.9% of the intervals can pass the K–S test and the $A^2$ test, respectively. For the other distributions, the Pareto distribution achieves the largest percentage of the 1-hour intervals that pass the K-S test over all UE clusters, i.e., up to 24.5%, while the Weibull distribution and the Tcplib distribution have at most 21.5% and 3.3% of intervals that pass the test.

## C SUPPLEMENTARY MICROSCOPIC ANALYSIS FOR MODEL VALIDATION

This appendix provides supplementary microscopic analysis (discussed in §4.5) to validate the proposed models. Specifically, for each device type, we plot the CDFs to examine the entire range of number of SRV_REQ/S1_CONN_REL per UE synthesized by our proposed method, and compare them with those generated by the baseline method.

Figure 11 shows that under Scenario 1 with 38K UEs, similar to phones, both connected cars and tablets have no explicit visual difference in the y-axis of the CDFs between the real and synthesized traces using our proposed traffic model. However, for the baseline, there exist visible differences in the y-axis of the CDFs, compared with the real trace. Those results confirm that our method achieves 3.25×~7.74×, 1.15×~2.65× and 2.80×~8.56× smaller maximum y-distance than the baseline method under Scenario 1 for phones, connected cars, and tablets, respectively.

As with Scenario 1, Figure 12 presents similar results under Scenario 2 with 380K UEs. Specifically, our proposed method has 3.52×~7.92×, 1.16×~3.63× and 3.07×~11.14× smaller maximum y-distance than the baseline method for phones, connected cars, and tablets, respectively.

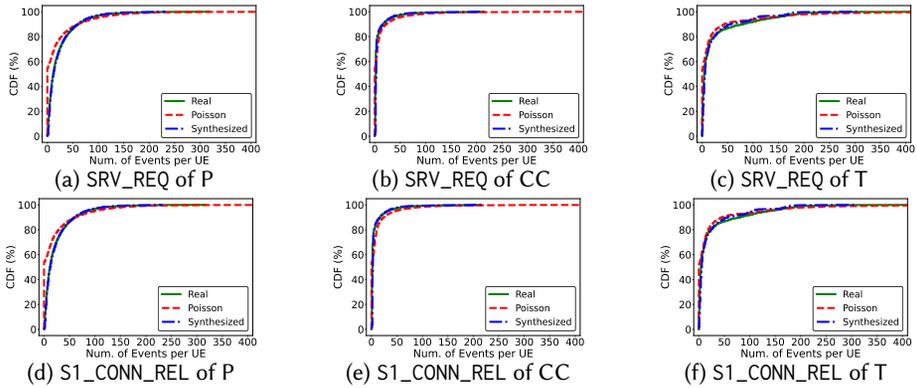

Fig. 11. Comparison of CDFs of number of SRV_REQ/S1_CONN_REL per UE between the synthesized and real 1-hour traces for three types of devices in Scenario 1 with 38,000 UEs.

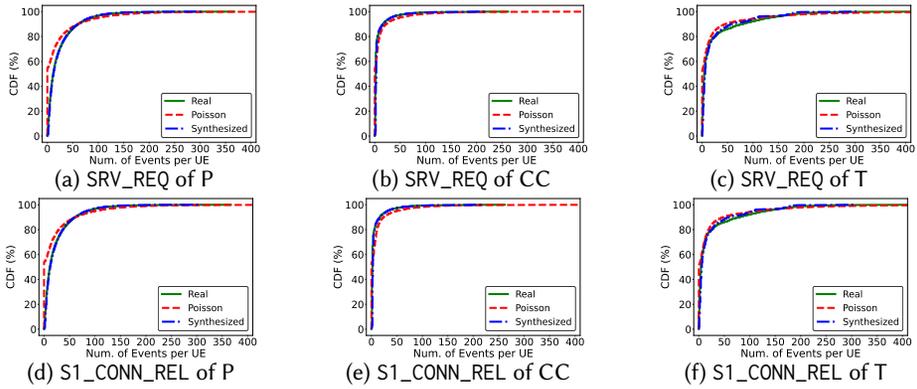

Fig. 12. Comparison of CDFs of number of SRV_REQ/S1_CONN_REL per UE between the synthesized and real 1-hour traces for three types of devices in Scenario 2 with 380,000 UEs.